\documentclass[twocolumn,superscriptaddress,showpacs,showkeys]{revtex4}

\usepackage{graphicx}
\usepackage[latin1]{inputenc}
\usepackage[english, swedish]{babel}
\usepackage{amssymb}

\def\lsim{\mathrel{\rlap{\lower4pt\hbox{\hskip1pt$\sim$}}
    \raise1pt\hbox{$<$}}}         
\def\gsim{\mathrel{\rlap{\lower4pt\hbox{\hskip1pt$\sim$}}
    \raise1pt\hbox{$>$}}}         

\newcommand{\ie}{{\it i.e.}}
\newcommand{\eg}{{\it e.g.}}

\newcommand{\be}{\begin{equation}}
\newcommand{\ee}{\end{equation}}
\newcommand{\ba}{\begin{eqnarray}}
\newcommand{\ea}{\end{eqnarray}}
\newcommand{\bt}{\begin{tabular}}
\newcommand{\et}{\end{tabular}}
\newcommand{\bfig}{\begin{figure}}
\newcommand{\efig}{\end{figure}}

\begin{document}

\selectlanguage{english}


\title{Strange quark asymmetry in the nucleon and the NuTeV anomaly}
\date{\today}
\author{J.~Alwall}
\email[E-mail: ]{johan.alwall@tsl.uu.se}
\affiliation{High Energy Physics, Uppsala University, Box 535, S-75121 Uppsala, Sweden}
\author{G.~Ingelman}
\email[E-mail: ]{gunnar.ingelman@tsl.uu.se}
\affiliation{High Energy Physics, Uppsala University, Box 535, S-75121 Uppsala, Sweden}
\affiliation{Deutsches Elektronen-Synchrotron DESY, D-22603 Hamburg, Germany}

\begin{abstract}
The NuTeV anomaly of a non-universal value of the fundamental parameter $\sin^2\theta_W$ in the electroweak theory has been interpreted as an indication for new physics beyond the Standard Model. However, the observed quantity depends on a possible asymmetry in the momentum distributions of strange quarks and antiquarks in the nucleon. This asymmetry occurs naturally in a phenomenologically successful physical model for such parton distributions, which reduces the NuTeV result to only about two standard deviations from the Standard Model. 
\end{abstract}

\preprint{TSL/ISV number}

\maketitle

The Weinberg angle $\theta_W$ is a fundamental parameter in the electroweak theory \cite{Eidelman:wy}. It quantifies the `mixing' of the electromagnetic and weak forces resulting in the photon and $Z^0$ as physical quanta mediating these interactions. Within the theory, the value of $\theta_W$ should be universal for all possible observables. It is, therefore, of great interest that the value of $\sin^2\theta_W$ recently extracted \cite{Goncharov:2001qe, Zeller:2001hh} from measurements of neutral and charged current neutrino and anti-neutrino cross-sections by the NuTeV collaboration is about three standard deviations above the value obtained from previous measurements, \eg\ from $e^+e^-$ annihilation at LEP. A number of possible explanations \cite{Davidson:2001ji} for this discrepancy have been suggested in terms of extensions to the Standard Model, but also in terms of inadequate descriptions of effects within the Standard Model.

An explanation of the latter kind would be a difference in the momentum distributions of strange ($s$) and anti-strange ($\bar{s}$) quarks in the nucleon, since neutrinos and anti-neutrinos interact differently with $s$ and $\bar{s}$. Although such $s\bar{s}$ pairs occurring as quantum fluctuations must have equal numbers of $s$ and $\bar{s}$ to conserve the strangeness quantum number, their momentum distributions may be different. The experimental evidence for such an asymmetry is as yet inconclusive \cite{Barone:1999yv,Olness:2003wz,Zeller:2002du,Mason:2004yf}, but there are theoretical motivations for it \cite{hadronic-fluct}. In this paper we show that an asymmetric strange sea is a natural consequence of a physical model for parton momenta that reproduces experimental data on the proton structure, such as the structure function $F_2$ and the $\bar u-\bar d$ asymmetry. In our model, the significance of the NuTeV anomaly is reduced to about two standard deviations, leaving no strong hint for new physics beyond the Standard Model. 

In the NuTeV experiment, the value of $\sin^2\theta_W$ is extracted from neutral and charged current cross-sections of neutrinos and anti-neutrinos, using a procedure based on the Paschos-Wolfenstein relation \cite{Paschos:1972kj,Zeller:2001hh}
\ba\label{eq-Rminus}
R^- & = & \frac{\sigma(\nu_{\mu}N\to \nu_{\mu}X)-
            \sigma(\bar{\nu}_{\mu}N\to \bar{\nu}_{\mu}X)}
	   {\sigma(\nu_{\mu}N\to \mu^- X)-
            \sigma(\bar{\nu}_{\mu}N\to \mu^+X)}
\nonumber \\
   & = & \frac{R^\nu - rR^{\bar{\nu}}}{1-r} 
   = g_L^2 - g_R^2 = \frac{1}{2} - \frac{5}{9}\sin^4\theta_W 
\ea
where the neutral current quark couplings $g$ are given by $\sin^2\theta_W$. 
Here, $r = \sigma(\bar{\nu} N\to \ell^+ X)/\sigma(\nu N\to \ell^- X) \sim 1/2$ and 
\be \label{eq-Rnu}
R^{\nu (\bar{\nu})} = \frac{\sigma(\nu(\bar{\nu}) N\to \nu(\bar{\nu})X)}
{\sigma(\nu(\bar{\nu}) N\to \ell^-(\ell^+)X)} = g_L^2 + r^{-1}g_R^2
\ee 
are the primary observables of NuTeV. The small deviation from an isoscalar target by an excess of neutrons over protons is accounted for in the NuTeV analysis, but  exact isospin symmetry in neutron and proton quark distributions is assumed, \ie\ $u_p(x)=d_n(x)$, $d_p(x)=u_n(x)$ and similarly for $\bar{u},\bar{d}$. As usual, $x$ represents the fractional momentum of the quark in the nucleon. The result for $R^-$ in eq.~(\ref{eq-Rminus}) is, furthermore, based on symmetry in the strange and charm quark sea distributions, \ie\ $s(x)=\bar{s}(x)$ and $c(x)=\bar{c}(x)$. If these symmetries are not present in the nucleon, the NuTeV result includes an erroneous shift in the value of $\sin^2\theta_W$ \cite{Davidson:2001ji,Zeller:2002du}. 

For the part of the nucleon sea arising from gluon splittings  $g\to q\bar{q}$ in perturbative QCD, symmetry is expected in the distributions of quarks and antiquarks, \ie\ $q(x)=\bar{q}(x)$. Conventional parameterisations of quark momentum distributions assume this symmetry also for the $x$-distribution at the start of the perturbative QCD evolution. However, for these sea distributions arising from the non-perturbative dynamics of the bound state nucleon there may well be such asymmetries. We show that nucleon fluctuations into $|\Lambda K\rangle$, where the $s$ quark is in the heavier $\Lambda$ baryon and the $\bar{s}$ is in the lighter $K$ meson, give a harder momentum distribution for the $s$ than the $\bar{s}$ which thus affects the NuTeV analysis.

\begin{figure}[thb]
\includegraphics*[width=30mm]{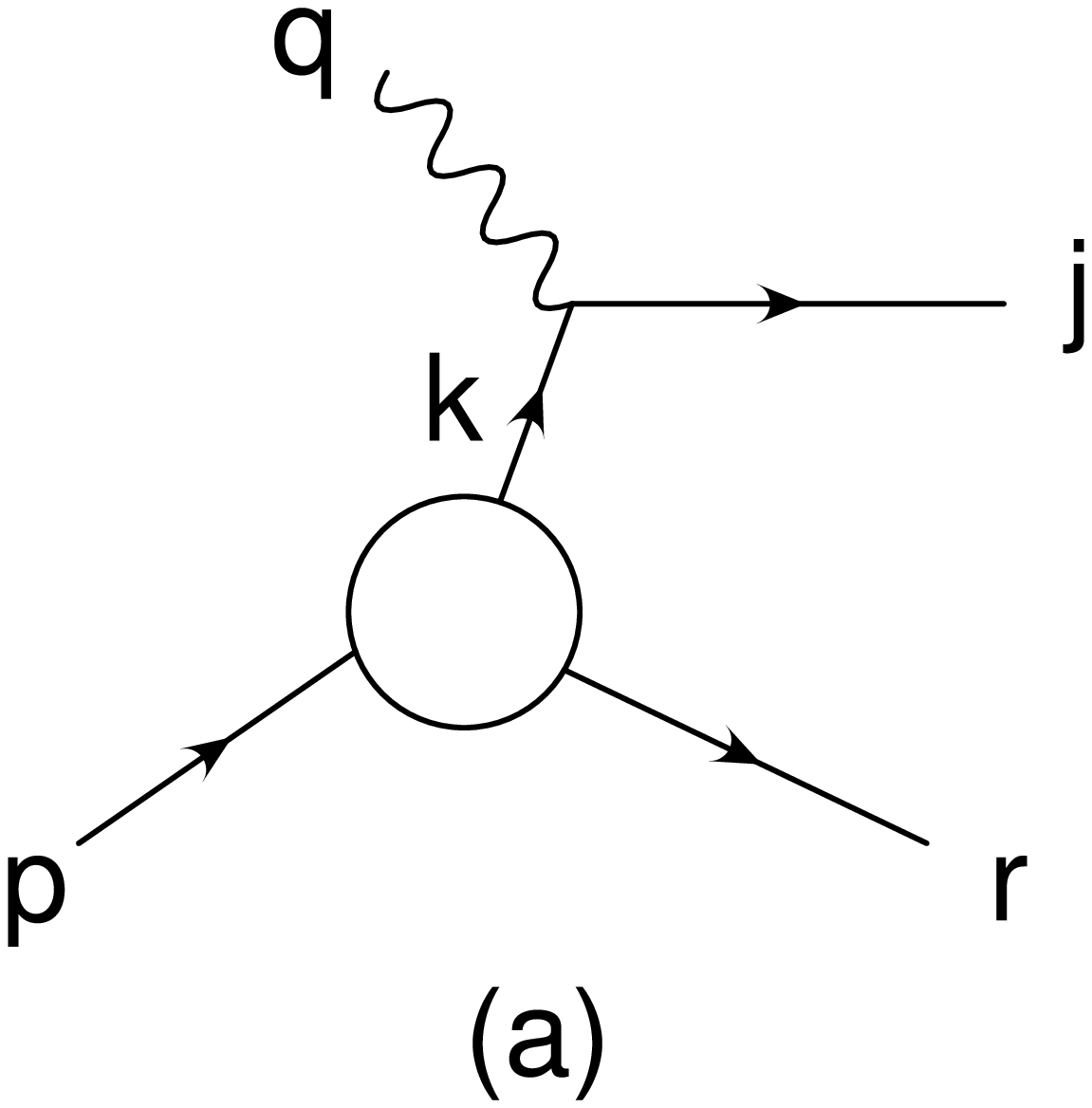}
\includegraphics*[width=30mm]{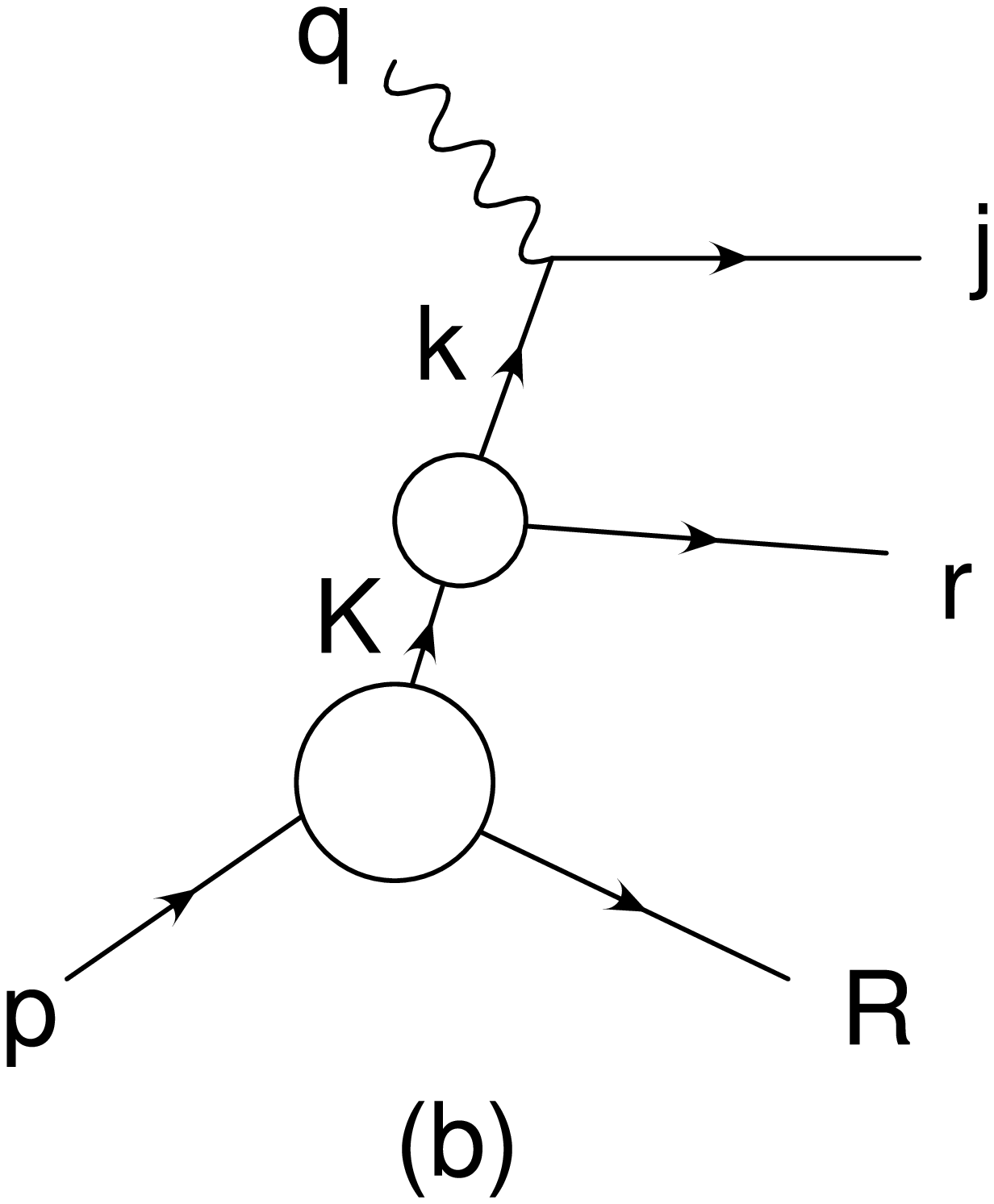}
\vspace*{-10mm}
\caption{\label{fig:fluct} Probing a valence parton in the proton and a sea parton in a hadronic fluctuation (letters are four-momenta).}
\end{figure}
We have previously \cite{Edin:1998dz,Edin:1999ep} presented a physical model giving the momentum distributions of partons in the nucleon, as illustrated in Fig.~\ref{fig:fluct}. More precisely, the model gives the $x$-shape of the parton distributions at a momentum transfer scale $Q_0^2\approx 1$~GeV$^2$, \ie\ $xq(x,Q_0^2)$ and $xg(x,Q_0^2)$, which provide an effective description of the non-perturbative dynamics of the bound state nucleon. Our approach gives the four-momentum $k$ of a single probed parton, whereas all other partons are treated collectively as a single remnant with four-momentum $r$ corresponding to integrating out all other information in the nucleon wave function. It is convenient to consider the nucleon rest frame where there is no preferred direction and hence the parton momentum distribution is spherically symmetric. The shape of the momentum distribution for a parton of type $i$ and mass $m_i$ is then taken as Gaussian
\be \label{eq-gaussian}
f_i(k) = N(\sigma_i,m_i) \exp\left\{-\textstyle\frac{(k_0-m_i)^2+k_x^2+k_y^2+k_z^2}{2\sigma_i^2}\right\}
\ee
which may be motivated as a result of the many interactions binding the parton in the nucleon. The width of the distribution should be of order hundred MeV from the Heisenberg uncertainty relation applied to the nucleon size, \ie\ 
$\sigma_i=1/d_N$. This Fermi motion inside the nucleon provides the `primordial transverse momentum', which has been extracted from deep inelastic scattering data and found to be well described by a Gaussian distribution of a few hundred MeV width \cite{primordial-kt} giving phenomenological support for this description. 

The energy component does not have the same simple connection to the Heisenberg uncertainty relation. It is assumed to have the same Gaussian fluctuation around the parton mass, such that partons can be off-shell at the soft scale of the binding interactions. This means a parton fluctuation life-time corresponding to the nucleon radius.

The momentum fraction $x$ of the parton is then defined as the light-cone fraction $x=k_+/p_+$. Here, four-momenta are expressed as $p=(p_+,p_-,\vec{p}_\perp)$ where the `plus' and `minus' components are $p_\pm=E\pm p_z$ and the $z$-axis defined by the probe. The fraction $x$ is then invariant under boosts along the $z$-axis and equivalent to the conventional momentum fraction $x=k_z/p_z$ in a frame where $p_z$ is large (`infinite momentum' frame). 

In order to obtain a kinematically allowed final state, one must impose the following constraints. The scattered parton must be on-shell or have a time-like virtuality (causing final state QCD radiation), \ie\ have a mass-squared in the range $m_i^2 \le j^2 < W^2$ ($W$ is the invariant mass of the hadronic system). Furthermore, the hadron remnant $r$ is obtained from energy-momentum conservation and must have a sufficient invariant mass to contain the remaining partons. These constraints also ensure that $0<x<1$. 

The parton distributions are obtained by integrating eq.~(\ref{eq-gaussian}) with these conditions. Using a Monte Carlo method this can be achieved numerically without approximations. (With further simplifying approximations it is possible to obtain an analytical expression for the parton densities \cite{Edin:1999ep}.) The normalisation of the valence distributions is provided by the sum rules 
\be \label{eq-flavoursumrule}
\int_0^1 dx\; u_v(x) = 2 \;\;\; {\rm and} \;\;\; \int_0^1 dx\; d_v(x) = 1 
\ee
to get the correct quantum numbers of the proton (and similarly for other hadrons). The gluon distribution is assumed to have the same basic Gaussian shape as the valence quarks, since they are all confined in the same region.
The gluon normalisation is given by the momentum sum rule, \ie\ $\sum_i \int_0^1 dx \; xf_i(x) = 1$, where the sum also includes sea partons that are generated as follows. 

Sea partons arise from the non-perturbative dynamics of the bound state nucleon, for which it is appropriate to use a hadronic quantum mechanical basis. Therefore we consider hadronic fluctuations, \eg\ for the proton 
\be \label{eq-hadronfluctuation}
|p\rangle = \alpha_0|p_0\rangle + \alpha_{p\pi}|p\pi^0\rangle + \alpha_{n\pi}|n\pi^+\rangle + \ldots + \alpha_{\Lambda K}|\Lambda K^+\rangle + \ldots 
\ee
Probing a parton $i$ in a hadron $H$ of such a fluctuation (Fig.~\ref{fig:fluct}b) gives a sea parton with light-cone fraction $x=x_H\, x_i$ of the target proton, \ie\ the sea distributions are obtained from a convolution of the momentum $K$ of the hadron and the momentum $k$ of the parton in that hadron. The momentum $\vec{K}$ of the probed hadron is given by a similar Gaussian as in eq.~(\ref{eq-gaussian}), with a separate width parameter $\sigma_H$, and the momentum $\vec{K}^\prime$ of the other hadron is then fixed by momentum conservation  in the nucleon rest frame. Both hadrons are taken on-shell, which fixes their energies. This implies that energy is not conserved at this intermediate stage, but is of course restored for the observable final state. With the hadron four-vectors specified one obtains the light-cone fraction $x_H=K_+/(K+K^\prime)_+$.

The above model for valence distributions is then applied to give the parton momentum in $H$ such that $x_i=k_+/K_+$ is obtained. The flavour sum rules in eq.~(\ref{eq-flavoursumrule}) must, of course, be modified to apply for $H$. The kinematical constraints to be applied in this case are $m_i^2\le j^2 < x_HW^2$ and that the remnants (see Fig.~\ref{fig:fluct}b) have invariant masses larger than their contained parton masses. A Monte Carlo method is used to simulate this two-step process by choosing $K$ and $k$, impose the constraints and obtain the momentum fraction $x$. By iterating the procedure the sea quark and gluon distributions are generated. 

The normalisation of the sea distributions is given by the amplitude coefficients $\alpha$. These are partly given by Clebsch-Gordan coefficients, but depend primarily on non-perturbative dynamics that cannot be calculated from first principles in QCD and are, therefore, taken as free parameters. 

\begin{figure}[htb]
\begin{center}
\includegraphics*[width=68mm]{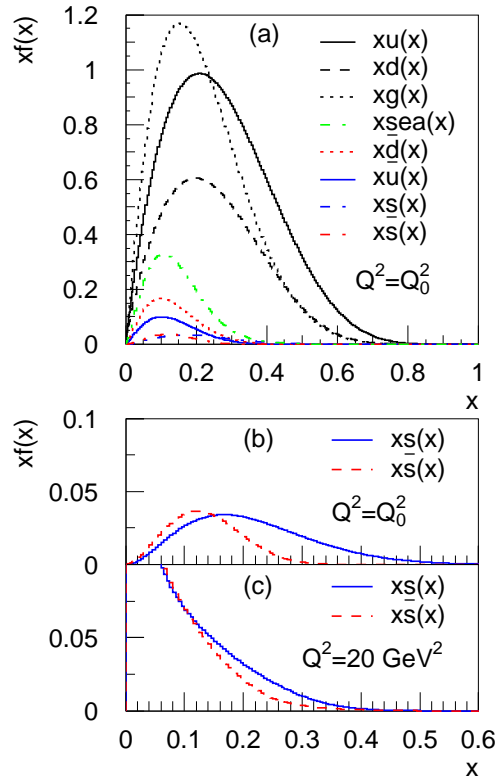}
\caption{\label{fig:pdfs}Parton momentum distributions $xf_i(x)$ for different parton species $i$ in the proton as obtained from the model with Gaussian momentum fluctuations in the proton giving valence distributions and in hadronic meson-baryon (\eg\ $\Lambda K^+$) fluctuations of the proton giving sea partons. (a) All partons and (b) enlarged plot of the strange quark ($s$) and antiquark ($\bar{s}$) sea at the scale $Q_0^2\sim 1$ GeV$^2$. (c) Strange quark and antiquark sea evolved with perturbative QCD to the scale $Q^2= 20$ GeV$^2$ representative for the NuTeV data. Note the harder momentum distribution of the $s$ quark (from $\Lambda$) relative to the $\bar{s}$ (from $K^+$).}
\end{center}
\end{figure}

This model provides valence and sea parton $x$-distributions as shown in Fig.~\ref{fig:pdfs}. These apply at a low scale $Q_0^2$, and the distributions at higher $Q^2$ are obtained by applying the next-to-leading order QCD evolution equations as implemented in \cite{Botje}. The proton structure function $F_2(x,Q^2)$ can then be calculated and the model parameters be fitted to data from deep inelastic scattering (DIS) resulting in the values:
\ba
\sigma_u &=& 180\, {\rm MeV},\; \sigma_d = 150\, {\rm MeV},\; \sigma_g = 135\, {\rm MeV}
\nonumber \\
\sigma_H &=& 100\, {\rm MeV},\;  \alpha_{sea}^2 = 0.06,\; Q_0^2 = 0.6\,  {\rm GeV}^2
\ea
where $\alpha_{sea}^2$ is the fraction of the proton momentum carried by sea quarks at the scale $Q_0^2$. This inclusive data can only be used to determine the overall normalisation $\alpha_{N\pi}^2$ for the dominating light quark sea from fluctuations with pions in eq.~(\ref{eq-hadronfluctuation}). As shown in \cite{Edin:1998dz}, the model reproduces the $F_2$ data well, which is remarkable in view of the model's simplicity with only six parameters.

The model also reproduces the observed asymmetry between the $\bar{u}$ and $\bar{d}$ distributions as a result of the suppression of fluctuations with a $\pi^-$ relative to those with a $\pi^+$, since the former require a heavier baryon (\eg\ $\Delta^{++}$) \cite{AI}. 

For the NuTeV data the most interesting fluctuation is $|\Lambda K\rangle$, where an asymmetry arise since the $s$ quark is in the heavier $\Lambda$ baryon and the $\bar{s}$ is in the lighter $K$ meson giving a harder momentum distribution for the $s$ than the $\bar{s}$. The same effect is present in fluctuations like $|\Sigma K\rangle$ and $|\Lambda K^\star\rangle$, which are all implicitly included in our generic $|\Lambda K\rangle$ fluctuation. Fluctuations where the $s\bar{s}$ pair is part of a meson wave function are here neglected since they are suppressed ($\phi$ due to large mass and $\eta$ due to Clebsch-Gordan coefficients).

The result of the model is shown in Fig.~\ref{fig:pdfs}b, which clearly shows a difference between the $xs(x)$ and $x\bar{s}(x)$ distributions as expected. Perturbative QCD evolution to larger $Q^2$ shifts these distributions to smaller $x$ and gives rise to a symmetric $s\bar{s}$ sea arising from $g\to s\bar{s}$, but the characteristic difference between $xs(x)$ and $x\bar{s}(x)$ is still present as shown in Fig.~\ref{fig:pdfs}c. 

\begin{figure}[thb]
\includegraphics*[width=90mm]{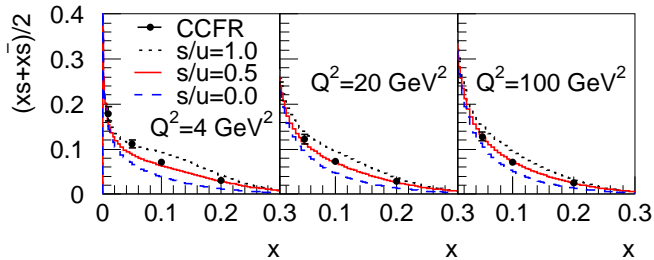}
\vspace*{-5mm}
\caption{\label{fig:CCFR} CCFR deep inelastic scattering data \cite{Bazarko:1994tt} on the strange sea distribution $(xs(x)+x\bar{s}(x))/2$ in the nucleon at different $Q^2$ compared to our model based on $|\Lambda K\rangle$ fluctuations (with normalisation `s/u'$=(s+\bar{s})/(\bar{u}+\bar{d})$ as discussed in the text) at a low scale and evolved to larger $Q^2$ with perturbative QCD that also adds a symmetric perturbative $s\bar{s}$ component.}
\end{figure}

The normalisation of the $|\Lambda K\rangle$ fluctuation cannot be safely calculated and is therefore taken as a free parameter which was fitted to data from the CCFR collaboration \cite{Bazarko:1994tt} on the strange sea $(s+\bar s)/2$, as shown in Fig.~\ref{fig:CCFR}. The data can be reproduced with an $s\bar{s}$ contribution from $|\Lambda K\rangle$ fluctuations with normalisation such that $\int_0^1dx(xs(x)+x\bar{s}(x))/\int_0^1dx(x\bar{u}(x)+x\bar{d}(x))\approx 0.5$, \ie\ the strange sea momentum fraction (at $Q_0^2$) is approximately half of that of a light sea quark, in agreement with the parton density analyses in \cite{Barone:1999yv,Olness:2003wz,Martin:1998sq}. This normalisation means that the coefficients $\alpha^2$ in eq.~(\ref{eq-hadronfluctuation}) scale essentially with the fluctuation time $\Delta t\sim 1/\Delta E$. Furthermore, this $s\bar{s}$ suppression is of similar magnitude as the ratio $P(s\bar{s})/P(u\bar{u})\approx 1/3$ of probabilities for quark-antiquark production in phenomenological hadronization models, such as the Lund model \cite{Andersson:ia}. Given the fact that both cases concern $s\bar{s}$ pair production in a non-perturbative process in a color field, this need not be surprising but indicate common features. 
\begin{figure}[thb]
\begin{center}
\includegraphics*[width=65mm]{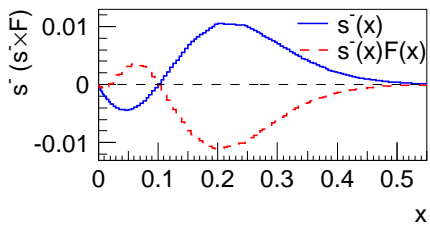}
\vspace*{-5mm}
\caption{\label{fig:s-asymmetry} The strange sea asymmetry $s^-(x) = xs(x)-x\bar{s}(x)$ (at $Q^2=20$ GeV$^2$) from the model and combined with the function $F(x)$ accounting for NuTeV's analysis giving $\Delta \sin^2\theta_W = \int_0^1 dx\, s^-(x) F(x) = -0.0017$.}
\end{center}
\end{figure}

With this strange sea we show in Fig.~\ref{fig:s-asymmetry} the resulting asymmetry $s^-(x) = xs(x)-x\bar{s}(x)$ and its combination with the folding function $F(x)$ provided by NuTeV \cite{Zeller:2002du} to account for their analysis and give the shift in the extracted value of $\sin^2\theta_W$. We obtain the integrated asymmetry $S^-=\int_0^1dx\, s^-(x)=0.00165$, and the shift $\Delta \sin^2\theta_W = \int_0^1 dx\, s^-(x) F(x)= -0.0017$. Thus, the NuTeV value $\sin^2\theta_W = 0.2277 \pm 0.0016$, which is about $3 \sigma$ above the Standard Model value $0.2227 \pm 0.0004$ \cite{Eidelman:wy}, would be shifted to 0.2260 which is only $2.0 \sigma$ above the Standard Model value. 

We note that the study in \cite{Cao:2003ny}, based on exponentially
suppressed $\Lambda K$ fluctuations, gave an inconclusive result on
the sign of $S^-$, and a too small strange sea to affect the NuTeV
anomaly. In \cite{Ding:2004ht} a result similar to ours was
obtained. Unfortunately, these studies provide no comparisons with the
measured strange sea, making the significance of their results
difficult to assess. As shown recently \cite{Catani:2004nc} higher
order perturbative effects give a negative contribution to $S^-$,
although significantly smaller than our positive non-perturbative
effect.

The experimental situation is at present unclear. In \cite{Barone:1999yv}, a positive $S^-$ was favoured based on several earlier experiments, whereas  NuTeV \cite{Zeller:2002du,Mason:2004yf} obtains the opposite sign based on their own data. However, the global analysis in \cite{Olness:2003wz} of the $s-\bar s$ asymmetry, including the NuTeV data as well as the CCFR data and using a very general functional form for $s^-(x)$,  gives a best fit value for the asymmetry $S^-$ of the same magnitude and sign as ours.

We conclude that an asymmetry in the strange quark sea of the nucleon, which arises naturally in models where non-perturbative sea quark distributions originate from hadronic fluctuations of the nucleon, may reduce the NuTeV anomaly to a level which does not give a significant indication of physics beyond the Standard Model.

\end{document}